\documentclass[reqno,12pt,a4paper]{amsart}

\usepackage{mathrsfs}
\usepackage{amssymb}
\usepackage{amsfonts}
\usepackage{latexsym}
\usepackage{amsthm}

\usepackage{graphicx}
\def\lb{\label}

\newcommand{\er}[1]{\textrm{(\ref{#1})}}

\begin{document}

%%%%%%%%%% Some definitions %%%%%%%%%%

%%%%%%%% Equations, theorems %%%%%%%%%
\renewcommand{\theequation}{\arabic{section}.\arabic{equation}}
\theoremstyle{plain}
\newtheorem{theorem}{\bf Theorem}[section]
\newtheorem{lemma}[theorem]{\bf Lemma}
\newtheorem{corollary}[theorem]{\bf Corollary}
\newtheorem{proposition}[theorem]{\bf Proposition}
\newtheorem{definition}[theorem]{\bf Definition}
\newtheorem{remark}[theorem]{\it Remark}
%\theoremstyle{remark}
%\newtheorem{remark}[theorem]{\bf Remark}

%%%%% Alphabet %%%%%
\def\a{\alpha}  \def\cA{{\mathcal A}}     \def\bA{{\bf A}}  \def\mA{{\mathscr A}}
\def\b{\beta}   \def\cB{{\mathcal B}}     \def\bB{{\bf B}}  \def\mB{{\mathscr B}}
\def\g{\gamma}  \def\cC{{\mathcal C}}     \def\bC{{\bf C}}  \def\mC{{\mathscr C}}
\def\G{\Gamma}  \def\cD{{\mathcal D}}     \def\bD{{\bf D}}  \def\mD{{\mathscr D}}
\def\d{\delta}  \def\cE{{\mathcal E}}     \def\bE{{\bf E}}  \def\mE{{\mathscr E}}
\def\D{\Delta}  \def\cF{{\mathcal F}}     \def\bF{{\bf F}}  \def\mF{{\mathscr F}}
\def\c{\chi}    \def\cG{{\mathcal G}}     \def\bG{{\bf G}}  \def\mG{{\mathscr G}}
\def\z{\zeta}   \def\cH{{\mathcal H}}     \def\bH{{\bf H}}  \def\mH{{\mathscr H}}
\def\e{\eta}    \def\cI{{\mathcal I}}     \def\bI{{\bf I}}  \def\mI{{\mathscr I}}
\def\p{\psi}    \def\cJ{{\mathcal J}}     \def\bJ{{\bf J}}  \def\mJ{{\mathscr J}}
\def\vT{\Theta} \def\cK{{\mathcal K}}     \def\bK{{\bf K}}  \def\mK{{\mathscr K}}
\def\k{\kappa}  \def\cL{{\mathcal L}}     \def\bL{{\bf L}}  \def\mL{{\mathscr L}}
\def\l{\lambda} \def\cM{{\mathcal M}}     \def\bM{{\bf M}}  \def\mM{{\mathscr M}}
\def\L{\Lambda} \def\cN{{\mathcal N}}     \def\bN{{\bf N}}  \def\mN{{\mathscr N}}
\def\m{\mu}     \def\cO{{\mathcal O}}     \def\bO{{\bf O}}  \def\mO{{\mathscr O}}
\def\n{\nu}     \def\cP{{\mathcal P}}     \def\bP{{\bf P}}  \def\mP{{\mathscr P}}
\def\r{\rho}    \def\cQ{{\mathcal Q}}     \def\bQ{{\bf Q}}  \def\mQ{{\mathscr Q}}
\def\s{\sigma}  \def\cR{{\mathcal R}}     \def\bR{{\bf R}}  \def\mR{{\mathscr R}}
                \def\cS{{\mathcal S}}     \def\bS{{\bf S}}  \def\mS{{\mathscr S}}
\def\t{\tau}    \def\cT{{\mathcal T}}     \def\bT{{\bf T}}  \def\mT{{\mathscr T}}
\def\f{\phi}    \def\cU{{\mathcal U}}     \def\bU{{\bf U}}  \def\mU{{\mathscr U}}
\def\F{\Phi}    \def\cV{{\mathcal V}}     \def\bV{{\bf V}}  \def\mV{{\mathscr V}}
\def\P{\Psi}    \def\cW{{\mathcal W}}     \def\bW{{\bf W}}  \def\mW{{\mathscr W}}
\def\o{\omega}  \def\cX{{\mathcal X}}     \def\bX{{\bf X}}  \def\mX{{\mathscr X}}
\def\x{\xi}     \def\cY{{\mathcal Y}}     \def\bY{{\bf Y}}  \def\mY{{\mathscr Y}}
\def\X{\Xi}     \def\cZ{{\mathcal Z}}     \def\bZ{{\bf Z}}  \def\mZ{{\mathscr Z}}
\def\O{\Omega}

\def\be{{\bf e}}   \def\bi{{\bf i}}
\def\bv{{\bf v}}

\def\mb{{\mathscr b}}
\def\mh{{\mathscr h}}
\def\me{{\mathscr e}}
\def\mk{{\mathscr k}}
\def\mz{{\mathscr z}}
\def\mx{{\mathscr x}}

\newcommand{\gA}{\mathfrak{A}}          \newcommand{\ga}{\mathfrak{a}}
\newcommand{\gB}{\mathfrak{B}}          \newcommand{\gb}{\mathfrak{b}}
\newcommand{\gC}{\mathfrak{C}}          \newcommand{\gc}{\mathfrak{c}}
\newcommand{\gD}{\mathfrak{D}}          \newcommand{\gd}{\mathfrak{d}}
\newcommand{\gE}{\mathfrak{E}}
\newcommand{\gF}{\mathfrak{F}}           \newcommand{\gf}{\mathfrak{f}}
\newcommand{\gG}{\mathfrak{G}}           %\newcommand{\gg}{\mathfrak{g}}
\newcommand{\gH}{\mathfrak{H}}           \newcommand{\gh}{\mathfrak{h}}
\newcommand{\gI}{\mathfrak{I}}           \newcommand{\gi}{\mathfrak{i}}
\newcommand{\gJ}{\mathfrak{J}}           \newcommand{\gj}{\mathfrak{j}}
\newcommand{\gK}{\mathfrak{K}}            \newcommand{\gk}{\mathfrak{k}}
\newcommand{\gL}{\mathfrak{L}}            \newcommand{\gl}{\mathfrak{l}}
\newcommand{\gM}{\mathfrak{M}}            \newcommand{\gm}{\mathfrak{m}}
\newcommand{\gN}{\mathfrak{N}}            \newcommand{\gn}{\mathfrak{n}}
\newcommand{\gO}{\mathfrak{O}}
\newcommand{\gP}{\mathfrak{P}}             \newcommand{\gp}{\mathfrak{p}}
\newcommand{\gQ}{\mathfrak{Q}}             \newcommand{\gq}{\mathfrak{q}}
\newcommand{\gR}{\mathfrak{R}}             \newcommand{\gr}{\mathfrak{r}}
\newcommand{\gS}{\mathfrak{S}}              \newcommand{\gs}{\mathfrak{s}}
\newcommand{\gT}{\mathfrak{T}}             \newcommand{\gt}{\mathfrak{t}}
\newcommand{\gU}{\mathfrak{U}}             \newcommand{\gu}{\mathfrak{u}}
\newcommand{\gV}{\mathfrak{V}}             \newcommand{\gv}{\mathfrak{v}}
\newcommand{\gW}{\mathfrak{W}}             \newcommand{\gw}{\mathfrak{w}}
\newcommand{\gX}{\mathfrak{X}}               \newcommand{\gx}{\mathfrak{x}}
\newcommand{\gY}{\mathfrak{Y}}              \newcommand{\gy}{\mathfrak{y}}
\newcommand{\gZ}{\mathfrak{Z}}             \newcommand{\gz}{\mathfrak{z}}

\def\ve{\varepsilon} \def\vt{\vartheta} \def\vp{\varphi}
\def\vk{\varkappa}  \def\vr{\varrho} \def\vs{\varsigma}

\def\Z{{\mathbb Z}} \def\R{{\mathbb R}} \def\C{{\mathbb C}}  \def\K{{\mathbb K}}
\def\T{{\mathbb T}} \def\N{{\mathbb N}} \def\dD{{\mathbb D}} \def\S{{\mathbb S}}
\def\B{{\mathbb B}}    \def\dk{{\Bbbk}}

%%%%% Arrows %%%%%

\def\la{\leftarrow}              \def\ra{\rightarrow}     \def\Ra{\Rightarrow}
\def\ua{\uparrow}                \def\da{\downarrow}
\def\lra{\leftrightarrow}        \def\Lra{\Leftrightarrow}
\newcommand{\abs}[1]{\lvert#1\rvert}
\newcommand{\br}[1]{\left(#1\right)}

\def\lan{\langle} \def\ran{\rangle}

%%%%% Typography %%%%%

\def\lt{\biggl}                  \def\rt{\biggr}
\def\ol{\overline}               \def\wt{\widetilde}
\def\no{\noindent}

%%%%% Math signs %%%%%

\let\ge\geqslant                 \let\le\leqslant
\def\lan{\langle}                \def\ran{\rangle}
\def\/{\over}                    \def\iy{\infty}
\def\sm{\setminus}               \def\es{\emptyset}
\def\ss{\subset}                 \def\ts{\times}
\def\pa{\partial}                \def\os{\oplus}
\def\om{\ominus}                 \def\ev{\equiv}
\def\iint{\int\!\!\!\int}        \def\iintt{\mathop{\int\!\!\int\!\!\dots\!\!\int}\limits}
\def\el2{\ell^{\,2}}             \def\1{1\!\!1}
\def\sh{\sharp}
\def\wh{\widehat}
\def\bs{\backslash}
\def\na{\nabla}
%%%%% Math operations %%%%%

%\bigskip
%\medskip
%\smallskip

\def\sh{\mathop{\mathrm{sh}}\nolimits}
\def\all{\mathop{\mathrm{all}}\nolimits}
\def\Area{\mathop{\mathrm{Area}}\nolimits}
\def\arg{\mathop{\mathrm{arg}}\nolimits}
\def\const{\mathop{\mathrm{const}}\nolimits}
\def\det{\mathop{\mathrm{det}}\nolimits}
\def\diag{\mathop{\mathrm{diag}}\nolimits}
\def\diam{\mathop{\mathrm{diam}}\nolimits}
\def\dim{\mathop{\mathrm{dim}}\nolimits}
\def\dist{\mathop{\mathrm{dist}}\nolimits}
\def\Im{\mathop{\mathrm{Im}}\nolimits}
\def\Iso{\mathop{\mathrm{Iso}}\nolimits}
\def\Ker{\mathop{\mathrm{Ker}}\nolimits}
\def\Lip{\mathop{\mathrm{Lip}}\nolimits}
\def\rank{\mathop{\mathrm{rank}}\limits}
\def\Ran{\mathop{\mathrm{Ran}}\nolimits}
\def\Re{\mathop{\mathrm{Re}}\nolimits}
\def\Res{\mathop{\mathrm{Res}}\nolimits}
\def\res{\mathop{\mathrm{res}}\limits}
\def\sign{\mathop{\mathrm{sign}}\nolimits}
\def\span{\mathop{\mathrm{span}}\nolimits}
\def\supp{\mathop{\mathrm{supp}}\nolimits}
\def\Tr{\mathop{\mathrm{Tr}}\nolimits}
\def\BBox{\hspace{1mm}\vrule height6pt width5.5pt depth0pt \hspace{6pt}}
\def\where{\mathop{\mathrm{where}}\nolimits}
\def\as{\mathop{\mathrm{as}}\nolimits}
\def\ctg{\mathop{\mathrm{ctg}}\nolimits}

%%%%%%%%%%%%% specialities %%%%%%%%%%%%%%

\newcommand\nh[2]{\widehat{#1}\vphantom{#1}^{(#2)}}
%{{\mathop{#1}\limits^\wedge}\vphantom{#1}^{(#2)}}
\def\dia{\diamond}

\def\Oplus{\bigoplus\nolimits}

%%%%%%%%%%% End of definitions %%%%%%%%%%

%%%%% OLD OLD OLD

\def\qqq{\qquad}
\def\qq{\quad}
\let\ge\geqslant
\let\le\leqslant
\let\geq\geqslant
\let\leq\leqslant
\newcommand{\ca}{\begin{cases}}
\newcommand{\ac}{\end{cases}}
\newcommand{\ma}{\begin{pmatrix}}
\newcommand{\am}{\end{pmatrix}}
\renewcommand{\[}{\begin{equation}}
\renewcommand{\]}{\end{equation}}
\def\eq{\begin{equation}}
\def\qe{\end{equation}}
\def\[{\begin{equation}}
\def\bu{\bullet}

%\bigskip
%\medskip
%\smallskip

\title[{Discretization of inverse scattering on a half line}]
 {Discretization of inverse scattering on a half line}

\date{\today}
\author[Evgeny Korotyaev]{Evgeny L. Korotyaev}
\address{Department of Analysis, Saint-Petersburg State University,   Universitetskaya nab. 7/9, St.
Petersburg, 199034, Russia, \ korotyaev@gmail.com, \
e.korotyaev@spbu.ru}


\begin{abstract}

We solve inverse scattering problem for Schr\"odinger operators with
compactly supported potentials on the half line. We discretize
S-matrix: we take the value of the  S-matrix on some infinite
sequence of positive real numbers. Using this sequence obtained from
S-matrix we recover uniquely the potential by a new explicit
formula, without the Gelfand-Levitan-Marchenko equation.

\end{abstract}     \maketitle

%\noindent{\small \textbf{Keywords:} inverse scattering problem} \ \
\ \
%{\bf Preliminary version}a

%\noindent {\small \textbf{AMS Subject classification:} }

\

{\small  Deducated to  Sergei Kuksin (Paris and Moscow) on the
occasion of his 65-th birthday}

\

%\vskip 0.25cm
\section {\lb{Sec1} Introduction and main results}
\setcounter{equation}{0}

\subsection{Introduction}
We consider  Schr\"odinger operators $T_q y=-y''+q(x)y,\ \ y(0)=0$
on $L^2(\R_+)$. We assume that  the potential $q$  belongs to the
real classes $\cL^\a$ or $\cP^\a$ defined by
 $$
 \cL^\a=\Big\{f\in L_{real}^\a(\R_+): \supp f\ss [0,1]\Big\}, \qqq
 \cP^\a=\Big\{f\in \cL^\a: \sup\supp f=1\Big\},\  \a\ge 1.
 $$
It is well-known (see \cite{CS89}, \cite{F59}, \cite{M86} and
references therein) that the operator $T_q$ has only purely
absolutely continuous spectrum $[0,\iy)$ plus a finite number $m\ge
0$ of negative eigenvalues $ E_1<...<E_{m}<0. $ We introduce the
Jost solutions $f_+(x,k)$  of the  equation
\[
\lb{eq1} -f_+''+qf_+=k ^2f_+,\ \ \ (x,k)\in \R_+\ts \C\sm \{0\},
\]
under the conditions $ f_+(x,k)=e^{ixk},\ x\ge 1$, and the Jost
function $\p(k)=f_+(0,k)$.
 The Jost function $\p$ is entire and satisfies
\[
\lb{asj} \p(k)=1+ O(1/k)\qqq as \qq  |k|\to \iy,\qq k\in \ol\C_+,
\]
uniformly in $\arg k\in [0,\pi]$. The function $\p(k)$ has exactly
$m$ simple zeros in $\C_+$ given by
\[
\lb{kn} k_{j}=i|E_j|^{1\/2}\in i\R_+, \qq j\in \N_m=\{1,2,...,m\},
\]
possibly one simple  zero at 0,  and an infinite number of zeros
(so-called resonances)  $0\le |k_{m+1}|\le |k_{m+2}|\le ..... $ in
$\C_-$.  The S-matrix  is defined by
\[
\lb{defSm} S(k)={\ol \p(k)\/\p(k)}={
\p(-k)\/\p(k)}=e^{-i2\x(k)},\qqq
 \ \ \ k \in \R\sm \{0\},
\]
 where the function $\x(k)$ is the
phase shift. Recall  that the function $\x$ is odd and is continuous
on $\R\sm \{0\}$ and satisfies: $\x(k)=O(1/k)$ as $ k\to \pm \iy$
and $\x(\pm 0)=\mp \pi \big (m+{n_o(\p)\/2}\big)$. Here $n_o(f)$ is
the multiplicity of 0 as a zero of a function $f$ and note that
$n_o(\p)\le 1$.

There are a lot of results about an inverse scattering  problem: to
determine the potential $q$ using the given  phase shifts $\x(k)$
defined on all $k>0$,  see \cite{M86}, \cite{F59}, \cite{CS89},
\cite{Kr53} and references therein. Marchenko \cite{M86} proved that
the mapping $ q\to {\it spectral\ data}= \{\x+eigenvalues+ norming \
constants\}$ is a bijection between a specific class of potentials
and spectral data (see also Krein \cite{Kr53}, \cite{Kr54}). We
discuss the case of compactly supported potentials, which associated
with resonances.
 Zworski  obtained first results about the
characterization problem (Proposition 8, p.293, Corollary p.295 in
\cite{Z87}) on the real line. He gave necessary and sufficient
conditions for some function to be a transmission coefficient for
some compactly supported potential $q$. But unfortunately, it is
shown in \cite{K05} that this statement is not correct. Inverse
problems (characterization, recovering, uniqueness) for compactly
supported potentials  in terms of resonances were solved by
Korotyaev for a Schr\"odinger operator with a compactly supported
potential on the real line \cite{K05} and the half-line \cite{K04},
see also Zworski \cite{Z02}, Brown-Knowles-Weikard \cite{BKW03}
concerning the uniqueness.  Moreover, there are other results about
perturbations of the following model (unperturbed) potentials by
compactly supported potentials: step potentials \cite{C06}, periodic
potentials \cite{K11h},  and linear potentials (corresponding to
one-dimensional Stark operators) \cite{K17}. We mention also that
inverse resonance scattering (characterization, recovering,
uniqueness) were discussed by Korotyaev and Mokeev for a Dirac
operator and for canonical systems on the half-line \cite{KM20} and
the real line \cite{KM20L}.

%for compactly supported potentials  in terms of resonances were
%solved by Korotyaev for a Dirac operator with a compactly supported
%potential

We consider the following inverse problem: to recover the compactly
supported  potential when the phase shifts are given at some
increasing sequence of energy. We give the physical motivation. When
conducting an experiment (scattering of a particle by a potential)
the phase shift is measured in a discrete set of points of energy.
By this discrete set of phase shift values (+ proper eigenvalues and
norming constants) a potential is restored using numerical methods.
Our main goal is to solve the inverse scattering problem on the
half-line according to the values of the phase shift in a discrete
set of points only.

Introduce the set $\cJ_\a$ of all possible Jost functions for $q\in
\cP^\a$:

{\bf Definition J.} {\it By   $\cJ_\a, \a\in [1,\iy)$ we mean the
class of all entire functions $f$  having the form
\[
\lb{J1}
\begin{aligned}
\textstyle f(k )=1+{\hat F(k)-\hat F(0)\/ 2ik}, \qqq \ k\in \C,
\end{aligned}
\]
where  $\hat F(k)=\int_0^1 F(x)e^{2ixk }dx$ and $F\in\cP^\a$. The
zeros of $f$ satisfies:

\no i)  The set of zeros (counted with multiplicity) of $f$ is
symmetric with respect to the imaginary line,
 $f(k)\neq 0$ for any $k \in\R\sm \{0\}$ and
 $f$ has possibly a simple  zero at $k=0$.

\no ii) All zeros $k_1,..,k_{m}, m\ge 0$ of the function $f$ in
$\C_+$ are simple, belong to $i\R_+$ and if they are labeled by
$|k_1|>|k_2|>...>|k_m|>0$, then they satisfy
\[
\lb{J2}
\begin{aligned}
 (-1)^jf(-k_j) >0,\ \qqq \forall \  j\in \N_m:=\{1,2,...,m\},
\end{aligned}
\]
where \er{J2} is equivalent
 to  the following   condition:

\no $f(-k_j)\neq 0,\ \ j\in \N_m$ and the function $f$ has an odd
number $\geq 1$ of zeros on any interval $(-k_{j}, -k_{j+1}), j\in
\N_{m-1}$ and an even number $\geq 0$ of zeros on the interval
$(-k_{m},0]$.}

 Introduce the set $\cS_\a$ of all possible S-matrices:
$$
\textstyle
\cS_\a=\Big\{S(k)={\p(-k)\/\p(k)}, k\in \R:  \p\in \cJ_\a  \Big\},\qqq
\a\ge 1.
$$
We sometimes write $\p(k,q), k_{n} (q),...$ instead of $\p(k),
k_{n}, ...$ when several potentials are being dealt with. In
\cite{K04} it was shown that the mapping $q\to \p(k,q)$
 is the bijection between  $\cP^1$ and $\cJ_1$.  We improve
this  result, including the characterization.

\begin{theorem}
\lb{T1}
i) A mapping $ q\to  \p(k,q)$ is a bijection between
$\cP^\a$ and $\cJ_\a$ for any $ \a\in [1,2]$.  Moreover, there
exists an algorithm to recover the potential $q\in \cP^\a$
 in terms of its eigenvalues and resonances.

\no ii) Each mapping $S : \cP^\a\to \cS_\a, \a\in [1,2]$ given by $
q\to S(k,q)$ is a bijection.

\no iii) Let a sequence $r_n>0, n\ge 1$ satisfy
$|r_n-{\pi\/2}n|<{\pi\/8}$ for all $n\ge 1$. If
$S(r_n,q_1)=S(r_n,q_2)$ for some $q_1,q_2\in \cL^1$ and  all $n\in
\N$, then we have $q_1=q_2$.

\end{theorem}

{\bf Remarks.} 1)  In the inverse theory of Marchenko \cite{M86} one
needs the phase shift plus eigenvalues and norming  constants. In
iii) we need only $S(r_n,q_1), n\ge 1$.

2) In  Section 3 we have additional results about  the mapping $q\to
(S(r_n,q))_1^\iy$. In Theorem \ref{T2} we discuss the local
isomorphism and    an injection.

%\newpage

\subsection{Main results}
 Define sets  of potentials:
$$
\begin{aligned}
\cL_+^\a=\Big\{q\in \cL^\a: \p(k,q)\neq 0,\ \forall \ k\in
\C_+\Big\}, \qq \cP_+^\a=\cL_+^\a\cap \cP^\a, \qq \a\ge 1.
\end{aligned}
$$
Let $q_*\in \cL^\a$.  Consider a Jost function $\p_*=\p(k,q_*)$ with
zeros $k_j, j=1,..,m$. We define a new function $F=B(k)\p_* (k)$
using the Blaschke product $B$ given by
\[
\label{Bk} B(k)=\prod_{j=1}^{m} \frac{k+k_j}{ k-k_j},\qqq k\in \C_+,
\]
Results from \cite{K04} (see Theorem  \ref{TK04} below) give that
$F\in \cJ_1$ and  $F=\p(k,q)$ is the Jost function for an unique
potential $q\in \cL_+^1$. Note that there are more results about
zeros of Jost functions in \cite{K04}, \cite{K04s}, \cite{MSW10}.
Thus  we have two steps to solve the following inverse problem:

\no $\bu$  To recover the potential $q_*$, when $q\in \cL_+^1$ and
$B(k)$ are given and $\p(k,q_*), \p (k,q)\in \cJ_1$.

\no $\bu$ The inverse problem for functions $\p(\cdot,q)\in \cJ_1$,
when $q\in \cL_+^1$.

We consider {\bf the first step}: to recover $q_*$, when $q\in
\cL_+^1$ and all zeros $k_j, j=1,..,m$ of $\p_*$ in $\C_+$ are
given. Jost and Kohn \cite{JK53} adapted the Gelfand-Levitan method
\cite{GL51} to determine a potential $q_*$ from the function $q$ and
the zeros of $\p(k,q_*)$ in $\C_+$. We describe their results for
our case, when a potential $q\in \cP_+^1$ be given and for
simplicity we assume that $m=1$. For any $k_*=ir, r>0$ and any
constant $\gc>0$ the function $\p_*(k)=\p(k,q){k-k_*\/k+k_*}$ is a
Jost function for a potential $q_*\in L^1(\R_+)$ given by
\[
\lb{1c} q_*=q+q_o, \qq q_o=-(\log A)'',
\]
where the function $A_x=1+ \gc\int_0^{x} \vp^2(s,k_*,q)ds$ and
$\vp(x,k,q)$ is the solutions of the equation $-\vp''+q\vp=k^2\vp$
under the conditions: $\vp(0,k,q)=0$ and $\vp'(0,k,q)=1$. Note that
$ q_o, q_o'\in L^1(\R_+)$. Moreover, $ q_o, q_o'\in L^1(\R_+)$ and
$\gc=\int_0^\iy \vp^2(x,k_*,q_*)dy>0$ is the normalizing constant,
which  can be any positive number. It is important that, they
determine asymptotics:
\[
\lb{3c}
\begin{aligned}
q_o= -{2\/\gc}(2r)^5e^{-2rx}(1+o(1))\qq \as \qq x\to \iy.
\end{aligned}
\]

Recall results from \cite{K04}: {\it Let $q \in \cP^1$ and let
$\p_*=\p(k,q){k-k_*\/k+k_*}\in \cJ_1$ for some $k_* \in i\R_+$. Then
$\p_*=\p(k,q_*)$  is  the Jost function for a unique
 potential $q_*\in \cP^1$ .}
 In case we have $q_o\in \cL^1$.

Thus asymptotics  \er{3c} of Jost and Kohn \cite{JK53} is not
correct for specific cases.
 Then the identity \er{1c} of Jost and Kohn \cite{JK53} jointly with the remark
 from \cite{K04} solve the first step.
Note that if we take $\p\in \cJ_\a$  for some $q \in \cP^\a,
\a\in[1,2]$, and $\p_*\in \cJ_1$, then \er{1c} gives that $q_*\in
\cP^\a$. We describe more the first step  in Section 4.

We consider {\bf the second step}, which is the main and the more
important. Introduce the real space $\ell^2=\ell^2(\N)$ of all
sequences $f=(f_n)_{n=1}^\iy$ equipped with the norm $  \| f
\|^2=\sum _{n\ge 1} |f_n|^2 <\iy$.

Consider two Sturm-Liouville problems for $q\in L^1(0,1)$:
\[
\lb{p1}  -y''+qy=\l y, \ \ y(0)=y(1)=0,
\]
\[
\lb{p2}  -y''+qy=\l y, \ \ y(0)=y'(1)=0.
\]
Let $ \m_{n}$ and $\t_{n}, n\ge 1$ be eigenvalues of the problem
\er{p1} and \er{p2} respectively.
 It is well known that these eigenvalues are simple, interlace $\t_1<\m_1<\t_2<\m_2<....$
and satisfy:
\[
\lb{JK}
\begin{aligned}
\ca \t_n=\pi^2(n-{1\/2})^2+\s_o+\wt \t_n,\qqq n\ge 1,\qq  \wt \t_n
\to 0\
  \as \ n\to \iy
\\
\m_n=(\pi n)^2+\s_o+\wt \m_n,\qqq n\ge 1,\qq \wt \m_n \to 0\  \as \
n\to \iy\ac ,
\end{aligned}
\]
and $\s_o=\int_0^1qdx$, see \cite{MO75}. Moreover, if $q\in
L^2(0,1)$, then we have $(\wt \t_n)_1^\iy\in \ell^2$ and $(\wt
\m_n)_1^\iy\in \ell^2$. In our consideration we use the following
crucial fact:   $q\in \cL_+^2$ iff the first eigenvalue $\t_1\ge 0$
of \er{p2}, i.e.,
\[
\lb{q+t1} q\in \cL_+^2 \qqq \Leftrightarrow \qqq \t_1\ge 0,
\]
see e.g., \cite{K20}. We fix  model (unperturbed) sequences
(corresponding to potential $q=0$)  by
$$
 p^o=(p_n^o)_1^\iy,\qq
p_n^o={\pi\/2}n,\qqq  \g^o=(\g_n^o)_1^\iy,\qq  \g_n^o=(p_n^o)^2.
$$

\begin{proposition}
\lb{T3} Let $q\in \cL_+^1$. Then the function $k-\x(k)$, is strongly
increasing in $k\in [0,\iy)$ and $1-\x'>0$. Moreover,  for each
$n\ge 1$ the equation $k-\x(k)=p_n^o$ has a unique solution $p_n\ge
0$ such that
\[
\lb{apn1}
\textstyle
p_n=p_n^o+\x(p_n), \qq \x(p_n)={\s_o+\s_n\/2p_n^o},\qqq
 \qq \s_o=\int_0^1qdx,
\]
\[
\lb{apn2}  \m_n=p_{2n}^2, \qqq \t_n=p_{2n-1}^2\qqq  \forall \ n\ge
1,
\]
where $\m_n, \t_n$ are eigenvalues of the problem \er{p1} and
\er{p2} respectively.  If in addition $q\in \cL_+^2$, then we have
$(\s_n)_1^\iy\in \ell^2$.
\end{proposition}

For a potential $q\in \cL_+^2$ due to Proposition \ref{T3} we  have
an increasing sequence $0\le p_1<p_2<....$, where
$p_n=p_n^o+{\s_o+\s_n\/2p_n^o}$ and $(\s_n)_1^\iy\in \ell^2$. We
define the spectral data $\gS_0$ by
$$
\begin{aligned}
\gS_o=\rt\{   (\s_o,\s)\in \R\os \ell^2: \ 0\le p_1<p_2<p_3<...,\
p_n=p_n^o+{\s_0+\s_n\/2p_n^o}, \ n\ge 1, \ \s=(\s_n)_1^\iy\rt\}.
\end{aligned}
$$
Using asymptotics \er{apn1} with $\x(p_n)={\s_o+\s_n\/2p_n^o}$ we
define the {\bf phase mapping} $\f: \cL_+^2\to \gS_o$ by:
\[
\lb{f3}
\begin{aligned}
q\to \f=(\s_0, \s),\qq \s_o=q_o:=\int_0^1qdx, \qq \s=(\s_n)_1^\iy.
\end{aligned}
\]
We formulate our main result about the inverse problem.

\begin{theorem}\lb{T4}
The phase mapping $\f: \cL_+^2\to \gS_o$ is a bijection  between
$\cL_+^2$ and $\gS_o$. Moreover, for any $\f \in \gS_o$ the
corresponding potential $q\in \cL_{+}^2$ has the form:
\[
\lb{q1} q(x)=\s_o-2{d^2\/dx^2} \log \Big(\G_\g\det\O(x,p)\Big),\ \ \
x\in (0,1),
\]
where $\O(x,p), x\in (0,1)$ is the infinite matrix whose elements
$\O_{n,j}$ are given by
\[
\lb{q2}
\begin{aligned}
\O_{n,j}(x,p)={\g_n-\g_n^o \/ \g_n-\g_j^o} \lt\{\cos
\sqrt{\g_n}x+{(-1)^n-\cos 2\sqrt{\g_n}\/\sin 2\sqrt{\g_n}}\sin
\sqrt{\g_n} x, {\sin p_j^o x\/p_j^o}\rt\}_{w},
\\
\g_n=p_n^2-\s_o=\g_n^o+\wt\g_n,\  {\rm where}\qq
\g_n^o=(p_n^o)^2=(\pi n)^2/4,\qq \qq (\wt\g_n)_1^\iy\in \ell^2,
\end{aligned}
\]
where $\{u,v\}_{w}=uv'-u'v$  and
\[
\lb{q3} \G_\g=\prod_{j>n\ge 1}\lt({\g_n-\g_n^o \/\g_n-\g_j}\cdot
{\g_n^o-\g_j \/  \g_n^o-\g_j^o}\rt).
\]

\end{theorem}

{\bf Remarks.}  1) Here $\O-I$ is the trace class operator and
$\det\O(x,p)$ is well defined.  The proof is based on the recovering
identity for Sturm-Liouvill problem on the unit interval from
\cite{K19}, which was adopted from the case of even potentials, see
page 117 in \cite{PT87}.

2) Recovering similar to \er{q1} is well known due to Jost and Kohn
\cite{JK53}. In addition similar formulas are used for
reflectionless potentials on the line, see e.g., \cite{AS81},
\cite{DT79}, \cite{F59}, where only eigenvalues and norming
constants are used. In \er{q1} we use only the phase shift.

3) We would like to note that there is a paper \cite{ASU15}, where
the authors consider the inverse scattering problem for
Schr\"odinger operators $H_q$ for $q\in \cL^1$. They  also use the
GLM equation as in \cite{K04}. Unfortunately, they do not mention
the previous works about it such that \cite{BKW03},\cite{K04},
\cite{Z02}.

{\bf Example.} We compute $q$ with one parameter, when  $\g_n, n\ge
1$ have the form: the first $\g_1=p_1^2, 0<p_1<\pi $ is free and all
other $\g_n=(p_n^o)^2, n\ge 2$ are frozen. Let
$$
v=\sin {\pi\/2} x, \qq u=\cos \sqrt{\g_1}x-{1+\cos
2\sqrt{\g_1}\/\sin 2\sqrt{\g_1}}\sin \sqrt{\g_1} x=C\sin \sqrt{
\g_1}(1-x),
$$
where $C=\ctg {\g_1}$.  They satisfy
$$u''=-\g_1 u,\ v''=-(\pi/2)^2 v,\ w=uv'-u'v,\qqq w'=uv''-u''v=Euv, \
E=\g_1-\g_1^o.
$$
 Then from Theorem \ref{T4} we determine the corresponding potential
 $q$ by
$$
q=-2 (\ln w )'' =-2 \Big({w'\/w}\Big)'=-2{(w')^2-ww''\/w^2}.
 $$
In this case $\O$ is a scalar. If we take the potential $q$ with
$\gm$ free parameters $0<\g_1<....<\g_N<(p_{N+1}^o)^2$ and all other
$\g_n=(p_n^o)^2, n\ge N+1$ are frozen, then $\O$ is  the $N\ts N$
matrix.

\subsection{Smooth potentials}
In order to discuss smooth potentials we define  the Sobolev space
$W_\a$ and the class $W_\a^+$ by
$$
W_\a=\{q,  q^{(\a)}\in L^2(0,1)\},  \qq  W_\a^+=\{q\in \cL_+^2:
q|_{(0,1)}\in W_\a\},\qqq  \a\ge 0.
$$
 Recall results from \cite{MO75}: if $q\in W_\a$, then
eigenvalues $ \m_{n}$ and $\t_{n}, n\ge 1$ of the problem \er{p1}
and \er{p2}   have the asymptotics
\[
\lb{Mx1}
\begin{aligned}
\sqrt{
\m_n}=\sqrt{\m_n^o}+{\s_0\/2\sqrt{\m_n^o}}+{1\/\sqrt{\m_n^o}}\sum_{1\le
j\le {\a+1\/2}}{a_j\/(4\m_n^o)^{j}}
 +{u_n\/ n^{\a+1}},
\end{aligned}
\]
\[
\lb{Mx2}
\begin{aligned}
\sqrt{
\t_n}=\sqrt{\t_n^o}+{\s_0\/2\sqrt{\t_n^o}}+{1\/\sqrt{\t_n^o}}\sum_{1\le
j\le {\a+1\/2}}{b_j\/(4\t_n^o)^{j}}
 +{v_n\/ n^{\a+1}},
\end{aligned}
\]
for some real $a_j, b_j$ and $(u_n)_1^\iy, (v_n)_1^\iy\in \ell^2$.
If $q\in W_\a^+$, then due to  \er{apn2} the sequences
$\t_n=p_{2n-1}^2, \ \m_n=p_{2n}^2,  n\ge 1$ satisfy
\er{Mx1}-\er{Mx2}. We define the spectral data  $\gS_\a, \a\ge 1$
(similar to the case $\a=0$) for $q\in W_\a^+$ by
$$
\begin{aligned}
\gS_\a=\rt\{ (\s_0,a,b,\s)\in \R\os\R^{2\a}\os \ell^2   & :
a=(a_j)_1^\a, (b_j)_1^\a,\s= (\s_n)_1^\iy,\qq 0\le p_1<p_2<p_3<...,
\\
 p_n=p_n^o+{\s_o+\wt p_n\/2p_n^o}, \qqq  & \wt p_n=\sum_{2\le 2j\le
\a}{a_j\/(2p_j^o)^{2j}}+{\s_s\/(2p_j^o)^{\a+1}},
 \qq n=2s,\\
&  \wt p_n=\sum_{2\le 2j\le
\a}{b_j\/(2p_j^o)^{2j}}+{\s_s\/(2p_j^o)^{\a+1}}, \qq n=2s+1 \rt\}.
\end{aligned}
 $$
If $q\in W_\a^+$ for some $\a\ge 1$, then the sequence
$p_{2n-1}=\sqrt{\t_n},\ p_{2n}=\sqrt{\m_n}, n\ge 1$ has the
asymptotics \er{Mx1}, \er{Mx2}. Thus  we have the mapping from
$W_\a^+$ into $\gS_\a$ given by
\[
\f_\a: q\to \{\s_o,a,b,\s\},\qq \s_o=q_o.
\]

 \begin{corollary} \lb{T1x}
Each mapping $\f_\a: W_\a^+\to \gS_\a, \a\ge 1$ is a bijection
between $W_\a^+$ and $\gS_\a$.

\end{corollary}

\no The plan of this paper is as follows. In Sect. 2  we prove the
main results. In Sect. 3 we consider the mapping $q\to S(r_n,q),
n\in \N$ for some increasing sequence $r_n>0, n\ge 1$. In Sect. 4 we
discuss results of Jost and Kohn.

\section {\lb{Sec2} Proof of main Theorems}
\setcounter{equation}{0}

\subsection{Preliminaries }

We recall some well known facts about entire functions (see e.g.,
\cite{Ko88}). An entire function $f(z)$ is said to be of
$exponential$ $ type$  if there is a constant $A$ such that
$|f(z)|\le $ const. $e^{A |z|}$ everywhere. The function $f$ is said
to belong to the Cartwright class $\cE_{Cart},$ if $f(z)$ is entire,
of exponential type, and satisfies:
$$
\int _{\R}{\log (1+|f(x)|)dx\/ 1+x^2}<\iy  ,\qqq \  \r_+(f)=0,\ \ \
\r_-(f)=2,
$$
where the types $\r_{\pm}(f)$ are given by $ \r_{\pm}(f)=\lim
\sup_{y\to \iy} {\log |f(\pm iy)|\/y}$.  Let $f(z)$ belong to the
Cartwright class $\cE_{Cart}$ and denote by $(z_n)_{n=1}^{\iy} $ the
sequence of its zeros $\neq 0$ (counted with multiplicity),  so
arranged that $0<|z_1|\le |z_2|\le \dots$. Then  we have the
Hadamard factorization:
\[
\lb{E1} f(z)=f(0)e^{iz}\lim_{r\to +\iy}\prod_{|z_n|\le r}\Big(1-{z\/
z_n}\Big),
\]
uniformly in every bounded disk and
\[
\lb{E2}
  \sum_{z_n\ne 0} {|\Im z_n|\/|z_n|^2}<\infty .
\]
We discuss Jost functions. It is well known that  the Jost solution
$f_+(x,k)$ satisfies the equation
\[
\lb{fs2} f_+(x,k)=e^{ixk}-\int_x^1{\sin k(x-t)\/k}q(t)f_+(t,k)dt,\qq
x\in [0,1],
\]
for all $(k,q)\in \C\ts \cP^1$. Note that the Jost function
$\p(k)=f_+(0,k)$ is entire. The Jost function $\p(k)$ is real on the
imaginary line and then satisfies $ \ol \p(k)=\p(-\ol k)$ for all $
k\in\C$.

Introduce the solutions $\vp (x,k), \vt(x,k)$ of the equation
$-y''+qy=k^2y$ under the conditions: $\vp(0,k)=\vt'(0,k)=0$ and
$\vp'(0,k)=\vt(0,k)=1$. Note that the function $\vp$ has the form
\[
\lb{fS}
\begin{aligned}
\vp(x,k)= {f_+(0,k)\/2ik}\Big(f_+(x,k)S(k)-f_+(x,-k)\Big) \qqq
\forall \ (x,k)\in \R_+\ts\R.
\end{aligned}
\]
From \er{fS} we obtain the well known identity for any $k>0, x\ge
1$:
\[
\lb{vp1} \vp (x,k)={|\p(k)|\/k}\sin (kx-\x(k) ).
\]
  We need standard results about the Jost function  for
  $q\in \cP^1$ (see e.g. \cite{K04}, \cite{K05}).

\no  $\bu$  The Jost function is expressed in terms of the
fundamental solutions $\vp ,\vt $ for all $k\in \C$ by
\[
\lb{fs4} e^{-ik}f_+(0,k)=\vp'(1,k)-ik\vp(1,k),\qq e^{-ik}f_+'(0,k)=i
k\vt (1,k)-\vt'(1,k).
\]
$\bu$ Recall that $\hat q(k)=\int_0^1q(x)e^{2ixk }dx$. The Jost
function $\p(k)=f_+(0,k), q\in \cL^1_\C$ satisfies
\[
\begin{aligned}
\lb{2.3} |\p(k)-1|\le   \o e^{|v|-v +\o},\qqq \qq v=\Im k,
\end{aligned}
\]
\[
\lb{2.4}
\begin{aligned}
&  \p(\cdot)=1+\p_1+\p_2,\qqq \p_1(k)={\wh q(k)-\wh q(0)\/ 2ik},
\\
&   |\p_2(k)|\le  \o^2e^{|v|-v +\o},
\end{aligned}
\]
where $ |k|_1=\max\{1,|k|\}$ and $\o=\min \{\|q\|, {\|q\|\/|k|}\}$
and $\|q\|=\int_0^1 |q(x)|dx$.

 $\bu$ The Jost function satisfies the following asymptotics
\[
\lb{2.5x} \log \p(k)=-\p_1(k)+{O(1)\/k^2},
\]
\[
\lb{2.5} \x(k,q)={\hat q(0)-\hat q_{c}(k)\/ 2k}+{O(1)\/k^2},
\]
as $|k|\to \iy, k\in \ol\C_+$, and uniformly with respect to $\arg
k\in [0,\pi]$ and $\hat q_c(k)= \int_0^1q(x)\cos 2kx dx$.

Below we need the following result.

\begin{theorem}
\lb{TK04} i) The mapping $q\to \p(k,q)$ acting from  $\cP^1$ to
$\cJ_1$ is a bijection.

\no ii) Let $\wt \p=  B(k)\p (k)$, where $B(k)=\prod_{j=1}^{m}
\frac{k+k_j}{ k-k_j}$. Then $\wt \p$ is the Jost function $\p(k,\wt
q)$ for an unique potential $\wt q\in \cL^1$ and $\p(k,\wt q)$ has
not zeros in $\C_+$.

\no iii) Let $q\in \cP^1$. Then  $\vr_+(\p)=0$ and $\vr_-(\p)=2$ and
the Jost function $\p(k,q)$ is given by
\[
\lb{Ha} \p(k)=k^{n_o} \p^{({n_o})}(0)e^{ik }\lim_{r\to
\iy}\prod_{|k_n|\le r, k_n\neq 0} \Big(1-{k \/ k_n} \Big), \ \ \ k
\in \C ,   \ \ \ n_0(\p)\in\{0,1\},
\]
uniformly on compact subsets of $\C$ and the shift function $\x $
has the form
\[
\lb{1.12}
 \x(k)=-\pi \Big(m+{n_o\/2}\Big)+ \int_0^k
\x'(t)dt,\ \ \ \ \ \ \ \x'(k )=1+\sum_{n=1,k_n\neq 0}^\iy {\Im k_n\/
|k -k_n|^2} ,\ \ \ k\geq 0,
\]
where $n_o=n_o(\p)$  and norming constants  $\gn_j:=\int_0^\iy
f_+^2(x,k_j)dx$ satisfy
\[
\lb{1.13} \gn_j=-i{\p'(k_j)\/\p(-k_j)}\qq \forall  \ j\in \N_m.
\]
Moreover, we can recover  the potential $q$ in terms of its
eigenvalues and resonances.

iv) The following trace formula holds true:
\[
\lb{tr11} -2k\Tr \rt(R(k)-R_0(k)\rt)={n_o\/k}+i+\lim_{r\to \iy}
\sum_{k_n\ne 0, |k_n|<r} {1\/k -k_n},
\]
\end{theorem}

{\bf Proof} of i)-iii) was given in \cite{K04}. We show \er{tr11}.
Define the Fredholm determinant
$$
D(k)=\det (I+Y(k))), k\in \C_+,\qq  \where \qq
Y(k)=|q|^{1\/2}R_0(k^2)|q|^{1\/2}\sign q.
$$
 It is well known that
$Y(k)$ is the trace class operator and thus the determinant is well
defined. Recall the known fact, see e.g., \cite{GK69}.  Let  an
operator-valued function $\O :\mD\to \cB_1$ be analytic for some
domain $\mD\ss\C$ and $(I+\O (z))^{-1}\in \cB$ for any $z\in \mD$.
Then for the function $F(z)=\det (I+\O (z))$ we have
\[
\label{S2F'z}
 F'(z)= F(z)\rm Tr (I+\Omega (z))^{-1}\Omega '(z).
\]
Thus due to \er{S2F'z} and the identity  $R=R_0-RqR_0$ we obtain the
known identity
$$
{D'(k)\/D(k)}=2k\Tr(I+Y(k))^{-1}Y'(k)=-2k\Tr    (R(k^2)-R_0(k^2)).
$$
The Hadamard factorization \er{Ha} yields the derivative of the Jost
function
$$
{\p'(k)\/\p(k)}={n_o\/k}+i+\lim_{r\to \iy} \sum_{k_n\ne 0, |k_n|<r}
{1\/k -k_n}.
$$
Using a known result of Jost--Pais \cite{JP51}, that the Jost
function is a Fredholm determinant, $\p=D$, we obtain \er{tr11}.
 \BBox

{\bf Remark.} Due to  \er{E1} the series in \er{1.12} converges
absolutely.

\subsection{Proof of  main results} We are ready to prove main
theorems.

\no {\bf Proof of Theorem \ref{T1}.}  We omit the proof of i), since
it repeats the proof of Theorem \ref{TK04}.

ii) We will prove that the mapping $S: \cP^\a\to \cS_\a$ given by $q\to
S(k,q)$ is an injection. Let $\p_j(k)$ be the Jost function for
potentials $q_j, j=1,2$ and let $S_j={\p_j(-k)\/\p_j(k)}$ be the
S-matrix. We assume that $S_1=S_2$. We show that $q_1=q_2$. The
properties of the class $\cJ_\a$ give that the functions $\p_1$ and
$\p_2$ have the same zeros $\ne 0$. Note that
$n_o(\p_2)=n_o(\p_1)\le 1$, since the functions $\p_1, \p_2$ have
the same even number of zeros on the interval $(-k_{m},0]$. This and
\er{Ha}  yield $\p_1=\p_2$ and i) implies $q_1=q_2$.

We show that the mapping $S : \cP^\a\to \cS_\a$ is onto (the
characterization). Let $S={f(-k)\/f(k)}$ for some $f\in \cJ_\a$.
Then due to i) there exists a unique $q\in \cP^\a$ such that
$f=\p(k,q)$ and then we obtain
$S={f(-k)\/f(k)}={\p(-k,q)\/\p(k,q)}$.

iii) Let $\p_j$ be the Jost function for the potential $q_j, j=1,2$.
Define the function
$$
F(k)=\p_1(k)\p_2(-k)-\p_2(k)\p_1(-k),\qqq k\in \C.
$$
The entire function $F$ has the following properties:

1) $F\in L^2(\R)$, since $F(k)=O(1/k)$ as $k\to \pm \iy$.

2) $F(\pm r_n)=0$ for all $r_n, n\in \N$, and $F(0)=0$.

3) The function $F$ has a  type  $\le 2$.

Then Paley-Wiener Theorem (see p. 30 in \cite{Ko88})   gives
$F(k)=\int_{-2}^2g(x)e^{-2ikx}dx$ for some $g\in L^2(-2,2)$. Recall
the Kadets Theorem \cite{Ka64}: If a sequence of real numbers $z_n,
n\in \Z$ satisfies $\sup_n |z_n-n|<{1\/4}$, then $e^{iz_nt}$ is a
Riesz basis of $L^2(-\pi,\pi)$. Thus the Kadets Theorem and the
properties 1)-3)  yield $F=0$ and then
 $ {\p_1(-k)\/\p_1(k)}={\p_2(-k)\/\p_2(k)}$ for all
$k\in \C$. Then i) of theorem implies  that $q_1=q_2$. \BBox

In order to prove main results we discuss the properties of the
sequence $p_n,n\ge 1$.

{\bf Proof of Proposition \ref{T3}.} Let $q\in \cL^1_+$. Then from
\er{1.12} we obtain that $1-\x'(k)>0$ for all $k\ge 0$ and
$k-\x(k)={\pi\/2}n_o+\int_0^k(1-\x'(t))dt>0$. Asymptotics \er{2.5}
yields that $k-\x(k)\to +\iy$ as $k\to +\iy$. Thus for each $n\ge 1$
the equation $k-\x(k)=p_n^o$ has a unique solution $p_n\ge 0$ such
that due to \er{2.5} we have
\[
\lb{fn} p_n=p_n^o+{\s_o+\s_n\/2p_n^o},\qqq  \s_o=q_o=\int_0^1qdx \qq
{\rm and} \qq \s_n\to 0\qq \as \ n\to \iy.
\]
Due to \er{vp1} the function $\vp (1,k)={|\p(k)|\/k}\sin (k-\x(k) ),
k>0$ has the following zeros $p_{2n}, n\ge 1$. Thus each
$p_{2n}^2=\m_n, n\ge 1$ is the eigenvalues of the problem
$-y''+qy=\l y, y(0)=y(1)=0$.

The function $\vp '(1,k)={|\p(k)|}\cos (k-\x(k) ), k>0$ has the
following zeros $p_{2n-1}, n\ge 1$. Thus each $p_{2n-1}^2=\t_n, n\ge
1$ is the eigenvalues of the problem $-y''+qy=\l y, y(0)=y'(1)=0$.

Let $q\in \cL_+^2$. Then from \er{2.5} and \er{fn}  we have
$p_n=p_n^o+{\s_o+\s_n\/2p_n^o}$, where $(\s_n)_1^\iy\in \ell^2$.
\BBox

{\bf Proof of Theorems \ref{T4}.} Recall that the sequence $
\g^o=(\g_n^o)_1^\iy, \g_n^o={(\pi n)^2\/4}$. We define the domain
(spectral data) $\G=\Big\{ (\ve_n)_1^\iy\in \ell^2:
\g_1^o+\ve_1<\g_2^o+\ve_2<..... \Big\}\ss \ell^2$. Recall Corollary
\ref{T3} from \cite{K19}: {\it Let $\mH=\{q\in L^2(0,1):
\int_0^1qdx=0\}$.  For $q\in \mH $ we define the sequence
$\g=(\g_n)_1^\iy$ by $\g_{2n-1}=\t_n, \g_{2n}=\m_n, n\ge 1$. The
mapping $q\mapsto \g(q)$ from $\mH$ to $\G$ is a real-analytic
isomorphism between $\mH$ and $\G$. Moreover, the potential $q$ has
the form \er{q1}.}

Due to Proposition \ref{T3}  for each $q\in \cL^2_+$ there exist a
 sequence $p=(p_n)_1^\iy$ such that

$\bu$  $p_{2n}^2=\m_n, n\ge 1$ is the eigenvalues of the problem
$-y''+qy=\l y, y(0)=y(1)=0$.

$\bu$ $p_{2n-1}^2=\t_n, n\ge 1$ is the eigenvalues of the problem
$-y''+qy=\l y, y(0)=y'(1)=0$.

\no For the sequence $p_n=p_n^o+{\s_0+\s_n\/2p_n^o}$, where
$(\s_o,\s)\in \gS_o$  we define a new sequence $\g\in \G$ by
$$
\g=(\g_n)_1^\iy,\qq
  \g_n=p_n^2-\s_o=\g_n^o+\wt\g_n, \qqq
(\wt\g_n)_1^\iy\in \ell^2.
$$
Then using Corollary  \ref{T3} from \cite{K19} we obtain the proof
of Theorems \ref{T4}. \BBox

{\bf Proof of Corollary \ref{T1x}.} Due to Theorem \ref{T3} the
mapping $\f_0:\cL_+^2\to \gS_o$ is a bijection. Recall that
\er{apn2} implies $\t_n=p_{2n-1}^2, \ \m_n=p_{2n}^2$,  for all $
n\ge 1$. Recall results from \cite{MO75}:

{\it The potential $q\in W_\a$ for some $\a\ge 0$ iff the
eigenvalues $\t_n$ and $\m_n, n\ge 1$ (corresponding to the problem
\er{p1} and \er{p2}) have the following asymptotics estimates
\er{Mx1}-\er{Mx2}.}

Then from this result we deduce that the mapping $\f_\a: W_\a^+\to
\gS_\a$ is a bijection between $W_\a^+$ and $\gS_\a$ for any $\a\ge
0$.  \BBox

%\newpage

\subsection {Even potentials}
Recall that $\m_n=\m_n(q),n\ge 1$ are the Dirichlet eigenvalues of
the problem \er{p1}  and $\t_n=\t_n(q), n\ge 1$ are the  mixed
eigenvalues of the problem \er{p2} on the unite interval
corresponding to $q$ and they satisfy $\t_1<\m_1<\t_2<\m_2<...$. We
need to recall results from \cite{K20}, which will be crucial for
us:
\[
\lb{L+} q\in \cL_{+}^2 \qq  \Leftrightarrow \qq \t_1(q)\ge 0\qq
\Leftrightarrow \qq \m_1(q)> 0, \qq \vp'(1,0,q)\ge 0.
\]
We discuss the inverse problem for the  case of even potentials
$q\in \cL_{+,e}^2$, where the set of even potentials is defined by
$$
\begin{aligned}
 \cL_{+,e}^2=\rt\{q\in \cL_{+}^2 :q(x)=q(1-x), x\in(0,1)
\rt\} .
\end{aligned}
$$
The condition, when $q\in \cL_{+,e}^2$ are formulated in terms of
$\t_1\ge 0$. In Theorem \ref{T4} we have used $\t_n, \m_n, n\ge 1$
for all $n\in \N$  and the condition \er{L+} was very natural. But
for even potentials we plan to use only the Dirichlet eigenvalues
$\m_n, n\ge 1$. Thus we have to rewrite conditions $q\in
\cL_{+,e}^2$ only in terms of Dirichlet eigenvalues \er{epq3}.

\begin{lemma}
\lb{Tep}

 Let $q\in \cL_{+,e}^2$. Then the function $\vp'(1,\l)$ has the
form
\[
\lb{epq1} \vp'(1,\l)=\cos \sqrt{\l}+\sum_{n\ge 1}{(-1)^n-\cos
\sqrt{\m_n}\/(\l-\m_n)\dot \vp(1,\m_n)}\vp(1,\l),\qq \l\in \R,
\]
where $\dot u={\pa\/\pa \l}u$  and the series converges uniformly on
compact sets in $\C$, in particular,
\[
\lb{epq2} \vp'(1,0)=1-\sum_{n\ge 1}{1-\cos (\sqrt{\m_n}-\pi n)\/\m_n
(-1)^n\dot \vp(1,\m_n)}\vp(1,0),
\]
where $(-1)^n\dot \vp(1,\m_n)>0$ and $\vp(1,0)>0$. Moreover, we have
\[
\lb{epq3}
\begin{aligned}
 q\in \cL_{+,e}^2 \qq  \Leftrightarrow \qq  \m_1(q)> 0, \qq
 \sum_{n\ge 1}{1-\cos (\sqrt{\m_n}-\pi n)\/\m_n
(-1)^n\dot \vp(1,\m_n)}\vp(1,0)\le 1.
\end{aligned}
\]
\end{lemma}

{\bf Proof.} We need the interpolation formula from
McKean-Trubowitz's paper \cite{MT76} (see Theorem 2, p. 174): let
$F=\vp'(1,\l)-\cos \sqrt{\l}$, then the following identity
\[
\lb{MT1} F(\l)=\sum_{n\ge 1}{F(\m_n)\vp(1,\l)\/(\l-\m_n)\dot
\vp(1,\m_n)},\qq \l\in \C,
\]
holds true, where the series converges uniformly on compact sets in
$\C$. It is well known that if $q\in \cL_{+,e}^2$, then
$\vp'(1,\m_n)=(-1)^n$ for all $n\ge 1$, see e.g., \cite{GT84}, which
yields $F(\m_n)=(-1)^n-\cos \sqrt{\m_n}$ and \er{epq1}, \er{epq2}.
From \er{epq2}, \er{L+} we obtain \er{epq3}. \BBox

For $q\in \cL_+^2$ due to Proposition \ref{T3} we  have an
increasing sequence $(p_n)_1^\iy$ such that
$$
0\le p_1<p_2<...., where \qq p_n=p_n^o+\x(p_n),\qq
\x(p_n)={\s_o+\s_n\/2p_n^o},\qq (\s_n)_1^\iy\in \ell^2,
$$
and $\s_0=\int_0^1qdx$. We consider even potentials $q\in
\cL_{+,e}^2$, then we need less spectral data and we take only even
sequences $p_{2n}=\m_n, n\ge 1$. Moreover, in this case the
Dirichlet eigenvalues $\m_n$ have to satisfy additional condition
\er{epq3}. Thus using these conditions we define the spectral data
$\gS_{o,e}$ for even potentials  by
$$
\begin{aligned}
\gS_{o,e}=\ca   (\s_o,\s_e)\in \R\os \ell^2: \ 0<p_2<p_4<...,\
p_{2n}=p_{2n}^o+{\s_0+\s_{2n}\/2p_{2n}^o}, \ n\ge 1, \qq
\s_e=(\s_{2n})_1^\iy,
\\
{\rm the\ sequence}\ (p_{2n})_1^\iy\  {\rm satisties}\ \er{con1} \ac
\end{aligned}
$$
\[
\lb{con1} \sum_{n\ge 1}{(-1)^n-\cos p_{2n}\/\m_n v'(\m_n)}\le 1,\qq
{\rm where}\qq v(\l)=\prod_{j\ge 1} (1-{\l\/\m_j}), \qqq
\m_j=p_{2j}^2.
\]
The additional condition \er{con1}  for potentials from $q\in
\cL_{+,e}^2$ corresponds to the fact that the eigenvalue $\t_1\ge
0$, see \er{epq3}. Using asymptotics \er{apn1} with
$\x(p_{j})={\s_o+\s_{j}\/2p_{j}^o}$ for even $j=2n$, similar to the
phase mapping $\f: \cL_{+}^2\to \gS_{o}$ we define the {\bf even
phase mapping} $\f_e: \cL_{+,e}^2\to \gS_{o,e}$ by:
\[
\lb{f3e}
\begin{aligned}
q\to \f_e=(\s_o, \s_e),\qq \s_o=q_o=\int_0^1qdx, \qq
\s_e=(\s_{2n})_1^\iy.
\end{aligned}
\]
 We formulate our result about the inverse problem for even potentials.
For this case we use only Dirichlet eigenvalues $\m_n=p_{2n}^2$ and
do not use eigenvalues  $\t_n=p_{2n-1}^2,n\in \N$
 for mixed boundary conditions. In order to recover the potential
 $q\in \cL_{+,e}^2$ in \er{dOqe} we need simple
modification of Dirichlet eigenvalues $\m_n=p_{2n}^2$ and we define
the sequences of shifted Dirichlet eigenvalues $\gm_n=\m_n-\s_o,
n\ge 1$, such that $ (\gm_n-(\pi n)^2)_1^\iy\in   \ell^2$.

\begin{theorem}\lb{T5e}
The even phase mapping $\f_e: \cL_{+,e}^2\to \gS_{o,e}$ is a
bijection between $\cL_{+,e}^2$ and $\gS_{o,e}$. Moreover, for any
$\f_e(q)=(\s_o,\s_e) \in \gS_{0,e}$ the corresponding potential
$q\in \cL_{+,e}^2$ has the form:
\[
\lb{dOqe}
\begin{aligned}
q(x)=\s_o-2{d^2\/dx^2} \log \lt(G_\gm\det\O(x,\gm)\rt),\ \ \ x\in
(0,1),
\\
\gm_n=\m_n-\s_o=(\pi n)^2+\wt\gm_n,\  {\rm where}\qqq
   (\wt\gm_n)_1^\iy\in   \ell^2,
\end{aligned}
\]
where $\O(x,\gm)$ is the infinite matrix whose elements $\O_{n,j}$
are given by
\[
\lb{dOe} \O_{n,j}(x,\gm)={\gm_n-(\pi n)^2 \/ \gm_n-(\pi j)^2}
\lt\{\cos \sqrt{\gm_n}x+{(-1)^n-\cos \sqrt{\gm_n}\/\sin
\sqrt{\gm_n}}\sin \sqrt{\gm_n} x, {\sin \pi j x\/\pi j}\rt\}_{w},
\]
where $\{u,v\}_{w}=uv'-u'v$  and $ G_\gm=\prod_{j>n\ge
1}\Big({\gm_n-(\pi n)^2 \/\gm_n-\gm_j}\cdot {(\pi n)^2-\gm_j \/ (\pi
n)^2-(\pi j)^2}\Big). $

\end{theorem}

{\bf Proof.} The proof for even potentials is similar to the generic
case of Theorem \ref{T4}, but the formulas \er{dOqe}, \er{dOe} for
even potentials are taken from p. 117 of \cite{PT87}. There is a
small difference between \er{q1} and \er{dOqe}.

Let $q\in \cL_{+,e}^2$. Then due to Theorem \ref{T4} and \er{epq3}
we deduce that $\f_e(q)=(\s_0,\s_e) \in \gS_{0,e}$.

Recall the results from \cite{PT87}. Define the space of even
potentials
$$
\begin{aligned}
\mH_e=\rt\{p\in L^2(0,1): \int_0^1pdx=0,\qq  p(x)=p(1-x)\  \forall \
x\in(0,1)\rt\}.
\end{aligned}
$$
Following to the book of P\"oschel and Trubowitz \cite{PT87} we
define the set $\gJ$  of all real, strictly increasing sequences by
$$
\gJ=\Big\{ s\!=\!(s_n)_1^\iy: s_1<s_2<....., \qq s_n\!=\!(\pi
n)^2\!+\wt s_{n\,},\qq \wt s=(\wt s_{n\,})_1^\iy\!\in\!\ell^2 \Big\}
$$
Let $\gm(p),n\ge 1$ be the Dirichlet eigenvalues for $p\in \mH_e$.
The mapping $p\to (\gm(p))_1^\iy$ acting from $\mH_e$ to the set
$\gJ$ is a real analytic isomorphism between $\mH_e$ and $\gJ$.
Moreover, the potential $p$ is recovered by the identity \er{dOqe}
($p:=q-\s_o$) in terms of its eigenvalues $(\gm(p))_1^\iy$.

Let $q\in \cL_{+,e}^2$. Then due to Theorem \ref{T4} the sequence
$\m_n=p_{2n}^2$ are the Dirichlet eigenvalues for $q\in \mH_e$. Due
to Lemma \ref{Tep} the sequence $(\s_o, \s_e)\in \gS_{o,e}$ and the
above results from \cite{PT87} give that the mapping $\f_e:
\cL_{+,e}^2\to \gS_{0,e}$ is an injection.

Let we have $(\s_o,\s_e) \in \gS_{o,e}$. Then we have an increasing
sequence $\m_n=p_{2n}^2, n\in \N$ and  the above results from
\cite{PT87} give that $\m_n$ are the Dirichlet eigenvalues for $q\in
\mH_e$ such that condition \er{con1} holds true, which jointly with
\er{epq3} gives $q\in \cL_{+,e}^2$. Thus the even phase mapping
$\f_e: \cL_{+,e}^2\to \gS_{0,e}$ is a bijection between
$\cL_{+,e}^2$ and $\gS_{0,e}$. Moreover, the above results from
\cite{PT87} gives \er{dOqe}. \BBox

%\newpage

\section {Analytic mapping }
\setcounter{equation}{0}

\subsection {Analytic properties of S-matrix}
Recall that $\vp (x,k)$ is the solutions  of the equation
$-\vp''+q\vp=k^2\vp$ under the conditions: $\vp(0,k)=0, \
\vp'(0,k)=1$.  We need properties of the fundamental solution
$\vp(x,k)=\vp(x,k,q)$, see e.g. \cite{PT87}.  Firstly we recall the
following standard estimates:
\[
\begin{aligned}
 \lb{Fu27}
& |\vp(x,k,q)|\le {e^{x|\Im k|+\o}\/|k|_1},\qqq
|\vp(x,k,q)-\vp_0(x,k)|\le {\o e^{x|\Im k|+\o}\/2|k|_1},
\\
& {\rm where}\qq |k|_1=\max\{1,|k|\},\qqq \o=\min \{\|q\|,
{\|q\|\/|k|}\},\qqq \|q\|=\int_0^1 |q(x)|dx,
\end{aligned}
\]
where $\vp_0(x,k)={\sin k x\/k}$, .
 Secondly we recall the well-known analytic properties.

\begin{lemma}
\lb{T31} The functions $\vp(1,k,q), \vp'(1,k,q)$  are entire on
$(k,q)\in \C\ts L_\C^2(0,1)$. Their gradients are given by
\[
\lb{af1} {\pa \vp(1,k,q)\/\pa q(x)}=(A \vp)(x,k,q),
\]
\[
\lb{af2} {\pa \vp'(1,k,q)\/\pa q(x)}=(B \vp)(x,k,q),
\]
where
\[
\lb{af5}
\begin{aligned}
A(x,\l,q)=\vp(1)\vt(x)-\vp(x)\vt(1), \qqq
B(x,\l,q)=\vp'(1)\vt(x)-\vp(x)\vt'(1).
\end{aligned}
\]
here  $\vp(x)=\vp(x,k,q), \vt (x)=\vt (x,k,q),....,$ for shortness.

\end{lemma}

We need properties of the Jost function as a function of a potential
from the Hilbert space.

  \begin{lemma}
\lb{T32} i) The function $\p(k,q)$ is entire on $\C \ts \cL_\C^2$
and its  gradient is given by
\[
\lb{dp1}
\begin{aligned}
{\pa \p(k,q)\/\pa q(x)}=\vp(x,k,q)f_+(x,k,q).
\end{aligned}
\]
ii) For each $q\in \cL_\C^2$ there exists $p\in \cL_\C^2$ such that
the following identities hold true:
\[
\lb{dp2} \p(k,q)=1+\hat g (k)=1+{\hat p(k)-\hat p(0)\/2ik}, \qqq
\forall\
 k\in\C,
\]
where the function $g(t)=\int_t^1p(x)dx$.
\end{lemma}
 {\bf Proof.} i) Lemma \ref{T31} and \er{fs4} imply that
 the function $\p(k,q)$ is  entire on $\C \ts \cL_\C^2$.
Let for shortness $f_+(x)=f_+(x,k,q),\  \vp(x)=\vp(x,k,q),...$ Using
\er{fs4} and Lemma \ref{T31} we obtain
$$
\begin{aligned}
{\pa \p(k)\/\pa q(x)}=e^{ik}\rt({\pa \vp '(1)\/\pa q(x)}-ik
{\pa\vp(1)\/\pa q(x)}\rt)
\\
=e^{ik}\vp(x)\lt(\rt[\vt (x)\vp'(1)-\vp(x)\vt'(1)\rt]-ik\rt[\vt
(x)\vp(1)-\vp(x)\vt(1)\rt]\lt )
\\
=e^{ik }\vp(x)\lt(\vt(x)\rt[\vp'(1)-ik \vp(1)\rt]+\vp
(x)\rt[-\vt'(1)+i k\vt (1)\rt]\lt)
\\
=\vp(x)\lt(\vt (x)\p (0)+\vp (x)\p' (0)\lt)=\vp(x)f_+(x).
\end{aligned}
$$
The statement ii) was proved in \cite{K04} for real $q$. The proof
for complex case is similar. \BBox

From these two lemmas we obtain properties of functions  $S(k,q)$.

  \begin{lemma}  \lb{T33}
   Each function $S(k,q)$ for any fixed $ k>0$ is real analytic  on $\cL^2$ and its gradient is given by
\[
\lb{gS1} {\pa S(k,q)\/\pa q(x)}=-2ik{\vp^2(x,k,q)\/\p^2(k,q)}.
\]
Moreover, let in addition, $b>0$. Then the following asymptotics
holds true:
\[
\lb{gS2}
\begin{aligned}
S(k,q)= 1+{q_0-\wh q_c(k)\/ik}+{O(\|q\|^2)\/k^2},\\
{\pa S(k,q)\/\pa q(x)}=-{2i\/k}\rt(\sin kx+{O(\|q\|)\/k}\rt)^2,
\end{aligned}
\]
as $|k|\to\iy, \Im k\in [-b,b]$, uniformly on the bounded subsets of
$[0,1] \ts \cL^2_\C.$

\end{lemma}
{\bf Proof.} Let for shortness $f_+(x,k)=f_+(x,k,q),\
\vp(x,k)=\vp(x,k,q),...$. Due to the definition $S(k)={
\p(-k)\/\p(k)}$ and using \er{fs4} and \er{dp1} \er{fS} we deduce
that
$$
\begin{aligned}
{\pa S(k)\/\pa q(x)}={1\/\p(k)}{\pa \p(-k)\/\pa q(x)}-
\cS_n(q){1\/\p(k)}{\pa \p(k)\/\pa q(x)}
\\
= \vp(x,-k){f_+(x,-k)\/\p(k)}-S(k)\vp(x,k){f_+(x,k)\/\p(k)}
\\
= {\vp(x,k)\/\p(k)}\rt(f_+(x,-k)-S(k)f_+(x,k)\rt)
=-2ik{\vp^2(x,k)\/\p^2(k)}.
\end{aligned}
$$

From \er{2.3}, \er{2.4} we obtain
$$
\begin{aligned}
& S(k,q)={\p(-k)\/\p(k)}=\rt(1+{q_0-\wh q(-k)\/2ik}+\p_2(-k)\rt)
\rt(1-{q_0-\wh q(k)\/2ik}+\p_2(k)\rt)^{-1}
\\
& =1+{q_0-\wh q_c(k)\/ik}+{O(\|q\|^2)\/k^2}.
\end{aligned}
$$
From \er{gS1}, \er{Fu27} and \er{2.3}, \er{2.4} we obtain
$$
\begin{aligned}
& {\pa S(k)\/\pa q(x)}=-2ik{\vp^2(x,k)\/\p^2(k)}= -{2i\/k}\rt(\sin
kx+{O(\|q\|)\/k}\rt)^2
\rt(1+{O(\|q\|)\/k}\rt)^{-2}\\
\end{aligned}
$$
which yields \er{gS2}. \BBox

\subsection{Inverse problem}
We consider the following inverse problem: to recover the potential
when the S-matrix $S(r,q)$  is given for some increasing sequence of
energy. Define the class of positive sequences $\cR=\{ (r_n)_1^\iy:
r_n>0,\qq  |r_n-{\pi\/2}n|<{\pi\/8} \qq  \forall \ n\ge 1\}$. For a
fixed sequence $(r_n)_{1}^\iy\in \cR$ we define the mapping $\P:
q\to (s_0,s)$ by
\[
\lb{defs} \P: q\to (s_0,s), \ \ s_0=q_0,\ \  s=(s_n)_1^\iy, \qq
s_n=q_0-ir_n\Big(S(r_n,q)-1\Big),\qq n\ge 1.
\]

  \begin{lemma}  \lb{T34}
Let $r\in \cR$. Then each $s_n(q)=q_0-ir_n(S(r_n,q)-1),\ n\ge 1$ is
real analytic  on $\cL^2$.  Moreover, its gradient is given by
\[
\lb{3x1} {\pa s_n(q)\/\pa q(x)}=1-2r_n^2{\vp^2(x,r_n)\/\p^2(r_n)},
\]
and  the following asymptotics hold true:
\[
\lb{3x2}
\begin{aligned}
s_n(q)= \wh q_c(r_n)+{O(\|q\|^2)\/n},\\
{\pa s_n(q)\/\pa q(x)}=\cos 2r_nx+{O(\|q\|)\/n},
\end{aligned}
\]
as $n\to\iy$, uniformly on the bounded subsets of $[0,1] \ts
\cL_\C^2$.
\end{lemma}
{\bf Proof.} Lemma \ref{T33} gives that each $s_n(q),\qq n\ge 1$ is
real analytic  on $\cL^2$. Using \er{gS1} we obtain \er{3x1}.
Substituting \er{gS2} into $s_n=q_0-ir_n(S(r_n,q)-1)$ we obtain
\er{3x2}.
 \BBox

\begin{lemma}\lb{TAe}
Let $q\in L^2(0,1)$. Then functions  $\vp^2(x,\sqrt{\m_n})$ and
$\vp^2(x,\sqrt{\t_n}), n\ge 1$ form the basis in $L^2(0,1)$.
\end{lemma}
{\bf Proof.} Define the space $ L_{ev}^2(0,2)=\big\{p\in L^2(0,2):
p(x)=p(2-x), x\in(0,2)\big\} $ of even functions. For $q\in
L^2(0,1)$  we define a even potential $\wt q\in
L_{ev}^2(0,2)$ by $\wt q(x)=\ca q(x), & x\in (0,1)\\
                  q(2-x), & x\in (1,2) \ac$.

We consider the Schr\"odinger operator $\wt H_D $ with the Dirichlet
boundary conditions on the interval $[0,2]$ given by $ \wt H_D
f=-f''+\wt q f,\  f(0)=f(2)=0$. Let $\wt \m_n$ and $\wt f_n, n\ge 1$
be eigenvalues and the corresponding eigenfunctions   of $ \wt H_D$.
The eigenvalues $\t_n, \m_n, n\ge 1$ and the corresponding
eigenfunctions $g_n=\vp(x,\sqrt{\t_n}), f_n=\vp^2(x,\sqrt{\m_n})$
for $q$ on $[0,1]$ satisfy
$$
\begin{aligned}
-g_n''+qg_n=\t_n(q)g_n, \qq  \textstyle   g_n(0)=g'_n(1)=0,
\\
-f_n''+qf_n=\m_nf_n, \qq  \textstyle   f_n(0)=f_n(1)=0.
\end{aligned}
$$
Direct calculations (see e.g. \cite{K19}) give that the
eigenfunctions $\wt f_n$ and the eigenvalues $\wt \m_n, n\ge 1$
satisfy $ \wt\m_{2n-1}=\t_{n}(q) $ and $ \wt\m_{2n}=\m_{n}(q)$ and
$$
f_{2n-1}(x)=\ca  g_n(x), & x\in(0,1) \\
         g_n(2-x), & x\in (1,2)\ac,
\qqq
         f_{2n}(x)=\ca  f_n(x), & x\in(0,1) \\
         -f_n(2-x), & x\in (1,2)\ac.
$$
  Recall that ${\pa \wt\m_n\/\pa \wt q}=f_n^2(x)$
  and the sequence of functions $f_n^2(x), n\ge 1$  forms a Riesz   basis in $L^2(0,1)$
  (see e.g. \cite{PT87}). Thus functions  $\vp^2(x,\sqrt{\m_n})$ and
$\vp^2(x,\sqrt{\t_n}), n\ge 1$ form a Riesz  basis in $L^2(0,1)$.
 \BBox

\begin{theorem}
\lb{T2}
 Let a sequence $(r_n)_1^\iy\in \cR$. Then

\no i) The mapping  $\P: q\to (s_0,s)$,  defined by \er{defs} acting
from $\cL^2$ into $\R\os \ell^2$ is real analytic and is  an
injection.

\no ii)  The mapping $\P'(0)={\pa \P(q)\/\pa q}|_{q=0}:\cL^2\to
\R\os\ell^2$ has the form
\[
\lb{ds1} \P'(0)=\F:\cL^2\to \R\os\ell^2
\]
and  is a linear isomorphism between $\cL^2$ and $\R\os\ell^2$,
where $\F$ is  the Fourier transformation $\F:\cL^2\to \R\os\ell^2$
defined by
\[
\lb{ds2} (\F h)_0=\int_0^1 h(x)dx,\qqq (\F h)_n=\int_0^1 h(x)\cos
2r_nxdx, n\ge 1.
\]

\no iii)  For each $q\in \cL^2$ the mapping ${\pa \P(q)\/\pa
q}:\cL^2\to \R\os\ell^2$ has an inverse.

\end{theorem}

\no {\bf Proof.} i) We recall the following result (see e.g.
\cite{PT87}):

{\it Let $f:\cD\to H$ be a mapping from an open subset $\cD$ of a
complex Hilbert space into a Hilbert space $H$ with orthogonal basis
$e_n, n\ge 1$. Then $f$ is analytic on $\cD$ if and only if $f$ is
locally bounded, and each coordinate function $f_n=(f,e_n):\cD\to
\C$ is analytic on $\cD$. Moreover, the derivative of $f$ is given
by the derivatives of its coordinate function:}
\[
\lb{TAbB2x} {\pa f\/\pa q}(h)=\sum_{n\ge 1}{\pa f_n\/\pa q}(h)e_n.
\]
Due to this result and Lemmas   \ref{T34}  the mapping $q\to \P(q)$
from $\cL^2$ to $\R\os\ell^2$ is real analytic. Theorem \ref{T1},
iii) yields  that this mapping is an injection.

ii) Consider the case $q=0$ and let $s_n'(q)={\pa s_n(q)\/\pa
q(x)}$. From \er{3x1} we obtain
\[
\lb{gs0} s_0'(q)=1,\qq s_n'(q)
\big|_{q=0}=1-2(r_n)^2{\vp^2(x,r_n,0)\/\p^2(r_n,0)}= \cos 2r_n x.
\]
This yields that $\P'(0)$ has the form \er{ds1}. The Kadets Theorem
(see the proof of Theorem \ref{T1}) imply that $1, \cos 2r_n x,\
n\ge 1$ forms  a Riesz  basis in $L^2(0,1)$.

iii) Let $q\in \cL^2$. Using the asymptotic estimate \er{3x2}, we
see that $\P' $ is the sum of the Fourier transform $\P'(0)=\F$ and
a compact operator $K$ for all $q\in \cL^2$. That is
$$
\P'(q)h=(h_0, (s_n'h)_1^\iy)=\F h +Kh,\qq h\in \cL^2,
$$
where $K$ is a compact operator. Consequently, $\P' $ is a Fredholm
operator. We show  now that the operator $\P' $ is invertible by
contradiction. Let $h\in \cL^2$ be a solution of the equation
\[
\lb{dsh} \P' (q)(h)=0,\ \ \ \  {\rm or} \ \ \ \ \ \
 \Big\{ (s_n'(q),h)=0, n\geq 0\Big\},
\]
for some fixed $q\in \cL^2$. In order to prove $h=0$ we  introduce
the function
$$
g(k)=k\int_0^1 h(x)\vp^2(x,k)dx, \ \ \ k\in \C,
$$
which is even and entire, where $\vp(x,k)=\vp(x,k,q)$. From
\er{dsh}, \er{3x1}  we deduce that
$$
g(\pm r_n)=\pm r_n\int_0^1 h(x)\vp^2(x,\pm r_n)dx=0,\qqq \forall \
n\ge 1.
$$
The entire function $g$ has the following properties:

1) $g\in L^2(\R)$, since $\vp(x,k)={\sin k x+O(1/k)\/k}$ as $k\to
\pm \iy$.

2) $g(\pm r_n)=0$ for all $r_n, n\in \N$, and $g(0)=0$.

3) The function $g$ has a  type  $\le 2$.

Then Paley-Wiener Theorem (see p. 30 in \cite{Ko88})   gives
$g(k)=\int_{-1}^1\wh g(x)e^{-2ikx}dx$. Then  the properties 1)-3)
and  the Kadets Theorem (see the proof of Theorem \ref{T1}) imply
that $g=0$.

 We need to show $h=0$. Due to Lemma \ref{TAe} the
sequence $\vp^2(x,\sqrt{\m_n})$ and $\vp^2(x,\sqrt{\t_n}), n\ge 1$
forms  a Riesz  basis in $L^2(0,1)$. Then $\int_0^1
h(x)\vp^2(x,k)dx=0$ for all $k\in \C$ we obtain $h=0$.
 \BBox

%\newpage

\section {\lb{Sec4} Jost-Kohn's  identities}
\setcounter{equation}{0}

We discuss the solution of Jost and Kohn \cite{JK53} for specific
cases. Our goal is to show that at some conditions on $k_*, q\in
\cP^1$ there exists a solution to the Gelfand-Levitan equation,
which gives the compactly supported function $q_*-q$ and the norming
constant $\gc$ is uniquely determined by $k_*, q$.

Let $\#(E)$ be the number of zeros of $\p$ (counted with
multiplicity) in the set $E\ss \C$.

{\bf Definition K.} {\it Let  $\p\in \cJ_1$. By   $\cK(\p)$ we mean
the class of all $k_*\in i\R_+ $ such that

i) $\p(-k_*)=0$ and $\p(k_*)\ne 0$.

ii) If  $k_*\in (k_1,+i\iy) $, then the number $\# (-k_*, -k_1)\ge
1$ is odd.

iii) If $k_*\in (k_j,k_{j+1})$ for some $j\in \N_{m-1}$, then $\#
(-k_j,-k_*)\ge 1$ is odd.

iv) If $k_*\in (0,k_m)$, then $\# (-k_m,-k_*)\ge 1$ is odd. }

Note that  $\# (-k_j,-k_{j+1})\ge 1$ can be any  odd number for a
Jost function $\p(k,q)$ with a specific potential $q$, see
\cite{K20}. We describe potentials $q_*$ corresponding the Jost
function $\p_*(k)=\p(k){k-k_*\/k+k_*}$. Note that the formula \er{1}
was obtained by Jost and Kohn \cite{JK53} for large class of
potentials. They obtained similar results for the case $\p_*(k)=
\p(k) B(k)$, where $ B(k)=\prod_{1}^{m}\frac{k+k_{*j}}{ k-k_{*j}},\
k\in \C_+$ is the Blaschke product and the norming constants
$\gc_1,...,\gc_m$. This result is discussed in many papers and
books,  see e.g., \cite{F59} and \cite{CS89}.

We determine $\vp(x,k)$ for $x>1, k=ir\in i\R_+$. We have
\[
\lb{Me9} \vp(x,k)=c_1e^{r(x-1)}+c_2e^{-r(x-1)},\qqq x>1,
\]
for some real $c_1, c_2$. Using \er{fs4} we obtain at $x=1$:
\[
\lb{Me10}
\begin{aligned}
\ca \vp(1,k)=c_1+c_2,
\\
\vp'(1,k)=-ikc_1+ikc_2 \ac, \qqq \ca c_1={ik\vp(1,k)-\vp_1'(1,k)
\/2ik}=-e^{-ik}{\p(k)\/2ik}
\\
c_2={ik\vp(1,k)+\vp_1'(1,k)  \/2ik}=e^{ik}{\p(-k)\/2ik} \ac.
\end{aligned}
\]

\begin{lemma}
\lb{TL1} Let $\p \in \cJ_1$ for some $q \in \cP^1$ and let
$\p(-k_*)=0$ at $k_*=ir, r>0$. Then
\[
\lb{M2} \vp(x,k_*)=-{\p(k_*)\/2r}e^{rx},\qqq \forall \ x\ge 1,
\]
\[
\lb{j2} \vp^2(1,k_*)-2r
\int_0^1\vp^2(t,k_*)dt=ie^{ik_*}\p'(-k_*)\vp(1,k_*).
 \]
i) If  in addition $k_*\in \cK(\p)$, then
$\p_*(k)=\p(k){k-k_*\/k+k_*}$ belongs to  $\cJ_1$ and
\[
\lb{j3} i\p'(-k_*)\vp(1,k_*)>0.
\]
ii) If  in addition $k_*\notin K(\p)$, then
$\p_*(k)=\p(k){k-k_*\/k+k_*}\notin \cJ$ and
\[
\lb{j4} i\p'(-k_*)\vp(1,k_*)<0.
\]

\end{lemma}
{\bf Proof.} From \er{Me10}, \er{fs4} and $\p(-k_*)=0$ we obtain
\er{M2}.

Differentiating the equation $-\vp''+q\vp=k^2 \vp$ with respect to
$k$ yields
$$
-\dot \vp''+q\dot \vp=2k \vp+k^2 \dot \vp.
$$
Multiplying this equation by $\vp$ and the equation $-\vp''+q\vp=\l
\vp$ by $\dot \vp$ and taking the difference, we obtain
$$
2k \vp^2=\vp''\dot \vp-\dot \vp''\dot \vp=\{\dot \vp,\vp\}',
$$
where $\{u,v\}=uv'-u'v$. Then we have the standard identity
\[
\lb{j1}
\begin{aligned}
2k\int_0^1\vp^2(t,k)dt=\{\dot \vp, \vp\}(t,k)\rt|_0^1= (\dot \vp
\vp'-\dot \vp' \vp)(1, k)=-{\vp} (ik\dot \vp +\dot \vp' )(1, k),
\end{aligned}
\]
since $\vp(0,k)=0$ and $\vp'(0,k)=1$ for all $k$ and
$\vp'(x,k)=-ik\vp(x,k)$ for $x\ge1$.

Due to $\p(-k_*)=0$ we obtain from \er{fs4} at $k=-k_*$:
\[
\lb{M6}
\begin{aligned}
&\big(e^{-ik}\p(k)\big)_k'=e^{-ik}\dot \p(k)=(\dot \vp'-i\vp-ik\dot
\vp )(1,k),
\\
& -e^{ik_*}\dot \p(-k_*)=\big(i\vp-\dot \vp'-ik_*\dot \vp
\big)(1,-k_*)= (i\vp+\dot \vp'+ik_*\dot \vp )(1,k_*),
\\
& i\vp(1,k)+e^{ik}\dot\p(-k)=-(ik\dot \vp +\dot \vp')(1, k),
\end{aligned}
\]
since $\vp(1,-k)=\vp(1,k)  $ for all $k$. Then from \er{j1}, \er{M6}
we obtain at $k=k_*$:
\[
\lb{Me17}
\begin{aligned}
2r \int_0^1\vp^2(t,k)dt={(i\vp(1,k)+e^{ik}\dot\p(-k))\vp(1,k)\/i}=
\vp^2(1,k)-ie^{ik}\dot\p(-k)\vp(1,k),
\end{aligned}
\]
which yields \er{j2}.

i)  Recall the needed result from \cite{K04}: {\it Let  $\p\in
\cJ_1$ and let $\p(z)=0$ for some $z\in i\R$. Then the function
$\P=B\p, B(k)={k+z\/k-z}$ is entire. Assume that the zeros of $\P$
satisfy i) and ii) in Definition J. Then the function $\P\in
\cJ_1$.} Using this result and our conditions on the $\p\in \cJ_1$
and $k_*\in i\R_+$ we deduce that $\p_*\in \cJ_1$.

We show \er{j3} for $k_*\in (k_j,k_{j+1})\ss i\R_+$ for some $j\in
\N_{m-1}$, where $\# (-k_j,-k_*)\ge 1$ is odd. The proof of other
cases is similar. Differentiating $\p(k)=\p_*(k){k+k_*\/k-k_*}$ at
$k=-k_*$ we have
\[
\lb{Mp} i\dot \p(k)={i\p_*(k)\/k-k_*}={i\p_*(k)\/-2k_*}
={\p_*(k)\/-2|k_*|}={(-1)^{j+1}\p_*(-k_*)\/(-1)^{j}2|k_*|}, \qq
(-1)^{j+1}\p_*(-k_*)>0.
\]
We need the following fact from \cite{K20}: if  $k_*\in
(k_j,k_{j+1})$, then  $k_*^2\in (\m_j,\m_{j+1})$, which yields
$(-1)^{j}\vp(1, k_*)>0$. Here $(\m_n)_{n\in \N}$ is an increasing
sequence of eigenvalues of the problem  $-y''+qy=\l y, y(0)=y(1)=0$
on the unit interval $[0,1]$. This jointly with \er{Mp} gives
\er{j3}.

 Similar arguments imply ii) and, in particular, \er{j4}. \BBox

\begin{theorem}
\lb{T1z}
 Let $\p\in \cJ_1 $  for some $q \in \cP^1$ and let
$k_* \in i\R_+ \sm \{k_1,..,k_m\}$ and a norming constant $\gc>0$.
Then
 $\p_*=\p(k){k-k_*\/k+k_*}$ is  the Jost function for
 a unique potential $q_*$ given by
\[
\lb{1} q_*=q+q_o, \qq q_o=-(\log A)'',
\]
where $A_x=1+ \gc\int_0^{x} \vp^2(s,k_*,q)ds$ and $q_o, q_o'\in
L^1(\R_+)$. Moreover, there are two cases:

i) Let $k_*\in \cK(\p)$. Then $\p_*\in \cJ$   and
$\gc_*={2|k_*|e^{|k_*|}\/i\p'(-k_*)\vp(1,
 k_*)}>0$. Moreover, we have $q_*\in \cP^1$  if $\gc=\gc_*$ and  $q_*$
is not  compactly supported if $\gc\neq \gc_*$.

 ii)  Let $k_*\notin \cK(\p)$. Then $\p_*\notin \cJ_1$   and  $q_*$ is not
 compactly supported for any $\gc>0$.
\end{theorem}

\no {\bf Proof.} Let $\p\in \cJ_1$  for some $q \in \cP^1$ and let
$k_* \in i\R_+ \sm \{k_1,..,k_m\}$. We recall the Jost-Kohn result
\cite{JK53}: for the function $\p_*(k)=\p(k){k-k_*\/k+k_*}$ and any
norming constant $\gc>0$
 there exists a unique potential $q_*$ given by \er{1}. In
order to determine the potential $q_*$ the Gel'fand-Levitan equation
is used, here the perturbed operator $T_*$ has the additional
eigenvalue $E_*=k_*^2$, see e.g. \cite{F59} or \cite{CS89}. The
Gel'fand-Levitan equation in this case is given by
\[
\lb{Me1}
\begin{aligned}
& G(x,t)=-G_0(x,t)-\int_0^{x} G_0(t,s)G(x,s)ds, \ \ \ t\geq x,
\\
& G_0(x,t)= \gc \vp(x)\vp(t),\qqq \vp(x)=\vp(x,k_*).
\end{aligned}
\]
This equation has a degenerate kernel and easily solved, since we
can rewrite $G$ in the form
\[
\lb{Me4} G(x,t)=g(x)\vp(t),
\]
where $g$ is an unknown function.  Then the Gel'fand-Levitan
equation becomes
\[
\lb{Me5}
\begin{aligned}
& \rt(g(x)+\gc\vp(x)+g(x)\gc\int_0^{x} \vp^2(s)ds\rt)\vp(t)=0,
\end{aligned}
\]
which has the solution $g(x)A_x =-\gc\vp(x), \qq A_x=1+\gc\int_0^{x}
\vp^2(s)ds$, and then
\[
\lb{Me8} g(x)=-\gc{\vp(x)\/A_x},\qqq  \qqq
G(x,t)=-\gc{\vp(x)\vp(t)\/A_x}.
\]
Thus the potential $q_*$ for the perturbed operator $T_{q_*}$  has
the form $q_*=q+q_o$, where $q_o$ is given by
\[
\lb{Me11} q_o=-(\log A)''={(A')^2\/A^2}-{A''\/A}={A'\/A^2}Q, \qq
Q=A'-{A''A\/A'},
\]
which yields \er{1} and it is the well-known Jost-Kohn
 identity (see \cite{JK53}).

i) Let $k_*\in \cK(\p)$ and let $a=-{\p(k_*)\/2r}$. Then from
\er{M2}, \er{Me11} we obtain for $x>1$:
\[
\lb{Me12}
\begin{aligned}
& \vp(x,k_*)=ae^{rx}, \qq  A'=\gc\vp^2(x), \qq A''=2r A',\qq
Q=A'-2rA,\qq Q'=0.
\end{aligned}
\]
The function $Q=A'-2rA$ is continuous on $\R_+$ and  a constant on
$x\ge 1$. Then due to \er{j2},
\[
\lb{Me13}
\begin{aligned}
 Q|_{x=1}=\gc \vp^2(1)-2rA_1
=\gc\rt(\vp^2(1)-2r\int_0^1\vp^2(t)dt\rt)-2r=\gc
ie^{ik}\dot\p(-k)\vp(1)-2r,
\end{aligned}
\]
since    $A_1=1+\gc \int_0^1\vp^2(t,k)dt$.  From \er{Me13} we deduce
that $Q=0$
 if we take $\gc={2re^{r}\/i\dot\p(-k_*)\vp(1, k_*)}$ and  then
$q_o(x)=0$ for all $x>1$.

ii)  Let $k_*\notin \cK(\p)$. Then Lemma \ref{TL1} implies that
$\p_*\notin \cJ_1$, since its zeros do not satisfy conditions i) and
ii) in Definition J. We show that $q_*$ is not
 compactly supported.   Assume that $q_*$ is
 compactly supported. Then due to Theorem A, the zeros of $\p_*$
 have to satisfy conditions i) and ii) in Definition J.
\BBox

%\newpage

\footnotesize\footnotesize

\no {\bf Acknowledgments.} \footnotesize   Our study was supported
by the RSF grant No 18-11-00032.

\end{document}